\documentstyle[preprint,prd,aps]{revtex}

\begin{document}
\draft
\title{\large \bf Why 3 + 1 metric rather than 4 + 0 or 2 + 2?}
\author{\bf H. van Dam and Y. Jack Ng\thanks{Corresponding author.
E-mail: yjng@physics.unc.edu}}
\address{Institute of Field Physics, Department of Physics
and Astronomy,\\
University of North Carolina, Chapel Hill, NC 27599-3255\\}
\maketitle
\bigskip

\begin{abstract}

Why does the physical 4-dimensional space have a 3 + 1 signature rather
than a 4 + 0 or a 2 + 2 for its metric?  We give a simple explanation based 
largely on a group-theoretic argument a la Wigner.  Applied 
to flat spaces of higher dimensions the same approach indicates 
that metrics with more than one time 
dimension are physically unacceptable because the corresponding irreducible 
unitary representations are infinite dimensional (besides the trivial 
representation).

\bigskip



\end{abstract}

\newpage

Many of us must have thought of the question why Nature
selects the Minkowskian metric for space-time.
\cite{HawPen,Niel1,Green,Teg,BranVa}  Is this
selection dictated by stability?\cite{Niel1}  Does the 3 + 1 metric 
have a dynamical
origin?\cite{Green}  Or is it possible that the anthropic principle 
allows only a metric with
1-time and 3-space dimensions?\cite{Teg}  Modern string theories make use
of more than four dimensions; it has been suggested that string winding
modes may yield a mechanism which allows at most three spatial dimensions
to become large.\cite{BranVa}  

Recently Borstnik and Nielsen\cite{Niel2} showed that, in any even
dimension $d$, only metrics with the signature corresponding to $q$ time +
$(d-q)$ space dimensions with odd $q$ exist.  This Letter is an elaboration
on the conclusion reached by them for the case $d=4$.  We show that it is only
natural for our World to have a 3 + 1 signature 
rather than a 4 + 0 or a 2 + 2 for its metric.  
For even or odd $d > 4$, our study indicates that only metrics 
with 1 time dimension are physically acceptable.  (Without loss 
of generality we take the number of time 
dimensions to be less than or equal to the space dimensions.)

Borstnik and Nielsen's work\cite{Niel2} is based on 
assuming certain equations of
motion (valid for all spins except for spin 0).  We take a more general
route following Wigner\cite{Wigner} who found all possible elementary forms
of quantum mechanics compatible with the inhomogeneous proper orthochronous
Lorentz group.  It is our opinion that Wigner's approach is to be
preferred as it finds all possible forms of quantum mechanics in terms of
irreducible representations.  The approach through equations and the
assumption of unitary representations of a group is not as general and not 
as clear cut.  We encountered such an example before in the discussion of
finite-spin tachyon equations in the context of Minkowski
space.\cite{vandam}  There we and Biedenharn used Wigner's approach to show
that the unitary irreps for tachyons can have either zero spin or
an infinite spin (infinite number of linearly independent states for a
given value of four momentum of the particle) and nothing else.  (See
below.)  In other
words, there are actually no tachyons of spin-$\frac{1}{2}$ etc even though 
(one may think) one can write down the Dirac equation or similar equations
for a particle with imaginary mass.

We start with a brief overview of Wigner's approach to find all
``elementary'' forms of quantum mechanics in Minkowski space, the group of
which we shall limit to the Poincare group (the inhomogeneous proper
orthochronous Lorentz transformations)\cite{Hout}.  One assumes that one
deals with a coherent Hilbert space where pure states are represented by a
ray {$e^{ia}\varphi$, for all real $a$.  The observables are transition
probabilities between pure states, which are given by the absolute value
squared of the inner product of the rays, $|(\varphi,\psi)|^2$.  One also
assumes that all these transition probabilities are observable.  An
invariance transformation is then a one to one mapping between rays which
preserves the absolute square of the inner products $(\varphi,\psi)$.
 
Wigner\cite{Wigner,Hout} started with a theorem which states that for any
one to one mapping between the rays,
\begin{equation}
{e^{ia}\varphi} \longleftrightarrow {e^{ia^{\prime}}\varphi^{\prime}},
\label{map}
\end{equation}
which is an invariance transformation, i.e.,
\begin{equation}
\left| (\varphi,\psi) \right| ^2 = \left| (\varphi^{\prime}, \psi^{\prime})
\right| ^2,
\label{inv}
\end{equation}
for all pairs of rays ${e^{ia}\varphi}$ and ${e^{ib}\psi}$, either one can
adjust all the phase factors, $e^{ia}$ etc, in such a way that one obains a
unitary mapping of the vectors
\begin{equation}
\varphi^{\prime} = U \varphi,
\label{uni}
\end{equation}
or it is possible to adjust the factors so that one obtains an antiunitary
mapping of the vectors
\begin{equation}
\varphi^{\prime} = U \varphi^{\ast},
\label{anti}
\end{equation}
where $\varphi^{\ast}$ is the complex conjugate of $\varphi$.

Since each continuous transformation of the Poincare group is a square of
another one, we may limit ourselves to Eq. (\ref{uni}).  However,
now we have a projective representation
\begin{equation}
U(P_{1} P_{2}) = e^{ia(P_{1},P_{2})} U(P_{1}) U(P_{2}).
\label{proj}
\end{equation}
Wigner showed in Ref.\cite{Wigner} that by going to the covering group of
the Poincare group, i.e., by replacing the homogeneous Lorentz group by its
covering group SL(2,C), one finally finds a true representation of the
covering group of the Poincare group.  This is important because, for one
thing, it gives us the $\frac{1}{2}$-integer spins.  This procedure
(of going to the covering group) removes the phase
factors; but it works only for some groups like the Poincare group or the
Euclidean group in three dimensions.

(Here the following side remark is perhaps relevant.  Wigner's approach is
indeed more
general than an outright assumption of a unitary representation in the
Hilbert space of vectors.  It is a nice exercise, for example, to convince
oneself that
there is no way to make the wave function of the Schrodinger equation
transform under a unitary representation of the inhomogeneous Galileo
group.
This is due to the fact that the phase factors of the rays cannot be
absorbed in a covering group; the
trouble here is caused by the boosts (the transformations which change
velocities).)

If one replaces the homogeneous Lorentz group in the Poincare group by
O(4), the rotations in a 4-dimensional space, the Poincare group
becomes the 4-dimensional Euclidean group.  The trick with the covering
group again gives a representation of that group.  The group space of the
covering group is a 4-sphere with its surface identified as one point, 
-$\mathbf{1}$.  That means it is the surface of a 5-sphere, which is simply
connected.

For O(2,2), we restrict ourselves to that part of the group the elements of
which may be continuously deformed into the unit element.  Again the phase
factors may be removed by going to the covering group of O(2,2).  Here it
is useful to note that the covering group of the complex Lorentz group,
which contains not only O(3,1) but also O(4) and O(2,2), is a direct
product SL(2,C) $\times$ SL(2,C),\cite{eight} where SL(2,C) is the covering
group of the real Lorentz group.  For O(2,2) we find a four-fold covering.
For the two commuting rotations $R_1$ and $R_2$ one
gets a covering $\pm U_1(R_1) \otimes \pm U_2(R_2)$; this comes from the covering group of
O(2,2), where $U_1(R_1)$ and $U_2(R_2)$ belong to SU(2).

Our task now is to find the building blocks of any quantum mechanics
(specifically the irreducible unitary representations) in the three groups:
(1) the quantum mechanical Poincare group for the case of (3 + 1)-metric;
(2) the quantum mechanical group with SL(2,C) replaced by the covering
    group of O(4) for the case of (4 + 0)-metric;
(3) the quantum mechanical group with SL(2,C) replaced by the covering 
    group of O(2,2) for the case of (2 + 2)-metric.
First one notices that these three groups are all semi-direct products with
the translation groups as normal subgroups.  To find the irreducible unitary
representations of such groups, one starts with diagonalizing the
commuting translation groups.  A translation by a 4-vector $a$ is then
represented by $U(a) = e^{i a \cdot p}$ with $p$ denoting the four momentum.

Let us first consider the O(4) case.  The O(4) transformations move $p$ all
over its mass sphere $p^2 = m^2$.  A subgroup O(3) leaves $p$ invariant and
it is called the little group\cite{Schur}.  To be more precise, we are
dealing with the covering group, so the little group is SU(2).  The
irreducible unitary representations of SU(2) are labelled by ``spin'' $0,
\frac{1}{2}, 1, \frac{3}{2}$, ...  For instance, the spin-1 irrep we get
acts on a three-component vector function $\varphi_{i}(p)$ with $p$
satisfying $p^2 = m^2$ and the inner product of two such wave functions is
given by
\begin{equation}
(\varphi,\psi) = \int d\mu(p) \sum_{i=1}^{3} \varphi_{i}^{\ast}(p)
\psi_{i}(p),
\label{inner}
\end{equation}
where $d\mu(p)$ is the rotationally invariant measure on the four sphere
$p^2 = m^2$.

Thus the possible elementary particles in 4-dimensional Euclidean geometry
are labelled by ``spin'' 0, $\frac{1}{2}$, 1, $\frac{3}{2}$, 2, ...  There are no
further irreducible unitary representations, therefore there is no place
for equations which describe, e.g., a spin-$\frac{1}{4}$ particle, if these
equations are to be invariant for the Euclidean group and if there is an
invariant inner product.  So the outcome of this simple exercise is that we cannot reject O(4) on grounds of 
spins (at least not from our group-theoretic point of view alone).\cite{anal}
But ultimately it is not physically acceptable since (unlike the O(3,1) case to be discussed presently) there are no photons,
there is no speed of light, etc.

Next we consider the O(3,1) case.  Here things are much more interesting as
the momentum vector $p$ can be time-like, light-like, or space-like.

For a time-like $p$, things are like the O(4) case, one has spin $0,
\frac{1}{2}, 1, \frac{3}{2}$, ... for a massive particle.

For a light-like $p$, the little group is noncompact; so at first sight its 
unitary representations must be infinite dimensional.  However, this group
contains two commuting ``translations''.  These may be eliminated by gauge
invariance.\cite{Kim}  The little groups are subgroups of the covering group.  Here
that covering group is SL(2,C), which covers the Lorentz group doubly.
Restricting the covering group to what remains of the little group (after
eliminating the two ``translations''), one gets $U(1)$ in the form
$e^{i\varphi / 2}$ where $\varphi$ is the angle of rotation.  Thus, for
$\varphi = 2\pi$ one gets $-1$, and one has both the integer as well as
half-integer helicities.

For a space-like $p$, i.e., the case of tachyons, the little group is the
two-fold covering group of a $2+1$ (O(2,1)) Lorentz group.  This group is
not compact.  Unlike the case of light-like $p$, there is no gauge
invariance.  Therefore the unitary irreps of this group are infinite
dimensional except for the trivial spin-0 representation.  Hence, tachyons
have either zero spin or infinite spin.\cite{vandam}  This is confirmed in
superstring models where the tachyons (in Minkowski space)
always have spin 0.\cite{Tye}    

Lastly let us consider the O(2,2) case.  Here one encounters the same
troubles as for the tachyons.  The little group (which is a subgroup of the
covering group of O(2,2)) is noncompact for any choice of $p$.

For a space-like or time-like $p$, the little group is O(2,1), i.e., 
noncompact, and is without a commuting normal subgroup.  So the ``spin'' 
is zero or infinite.

For a light-like $p$, again the little group does not contain an abelian
normal subgroup.  This is to be contrasted with the O(3,1) case, for which
an abelian normal subgroup allows the little group, via gauge invariance,
to be reduced to rotations around the direction of $\vec{p}$, yielding 
helicities 0, $\pm\frac{1}{2}, \pm 1$, ... via the covering group.
Thus, for O(2,2), the ``spin'' is either zero or infinite.

The very same approach applies to flat spaces of 
higher (even or odd) dimensions ($d > 4$).  
Again the translations are an invariant commuting subgroup of the group of 
transformations of space-time which are continuously connected 
to the unit element (no time inversions or spatial reflections).  One 
diagonalizes the translations, this gives a $d$-vector $p$.  The little group
is the subgroup of the covering group of O($d-q,q$) which leaves $p$ 
invaraint.  
One finds that, whenever the smaller 
of $d - q$ and $q$ is greater than one, 
one has only infinite ``spin'' or (via the trivial 
representation) ``spin'' zero, just as for the O(2,2) case.  Thus for higher 
dimensional spaces, $d > 4$, we have only 
``spin'' zero or infinite ``spin'' for the elementary particles unless 
one has O($d$) (not interesting as argued
above for the $d = 4$ case), or O($d-1,1$).

To summarize, we have shown, a la Wigner, why our 4-dimensional World has
a 3 + 1 signature rather than a 4 + 0 or a 2 + 2 for its metric.  A 
(4+0)-world has no interesting dynamics 
while a (2+2)-world can only have spin-0 particles.
But a (3+1)-world can be rich in both contents and dynamics, as befitting a
physical world --- like the one we live in.  
By the same criteria, of all the higher dimensional spaces only 
those metrics with one time dimension are physically acceptable.
We conclude that space-time can only have one time dimension.

\bigskip

\begin{center}
{\bf Acknowledgments}\\
\end{center}

This work was 
supported in part by the US Department of Energy under \#DE-FG05-85ER-40219, 
and the Bahnson Fund at the University of North Carolina at Chapel Hill.


\bigskip

\begin{references}

\bibitem{HawPen}
R. Penrose and W. Rindler, {\it Spinors and Space-time} (Cambridge
University Press, 1986), p.235; R. Bousso and S.W. Hawking, hep-th/9807148.

\bibitem{Niel1}
H. B. Nielsen and S. E. Rugh, hep-th/9407011.

\bibitem{Green}
J. Greensite, Phys. Lett. {\bf 300B} (1993) 34.

\bibitem{Teg}
M. Tegmark, Class. Quant. Grav. {\bf 14} (1997) L69.

\bibitem{BranVa}
R. Brandenberger and C. Vafa, Nucl. Phys. {\bf B316} (1989) 391.

\bibitem{Niel2}
N. M. Borstnik and H. B. Nielsen, Phys. Lett. {\bf 486B} (2000) 314 and
references therein.

\bibitem{Wigner}
E. P. Wigner, Ann. Math. {\bf 40} (1939) 149.

\bibitem{vandam}
H. van Dam, Y. J. Ng, and L. C. Biedenharn, Phys. Lett. {\bf 158B} (1985)
227.

\bibitem{Hout}
R. M. F. Houtappel, H. van Dam, and E. P. Wigner, Rev. Mod. Phys. {\bf 37}
(1965) 595; A. S. Wightman, in {\it Dispersion Relations and Elementary
Particles} (John Wiley, New York, 1960), ed. by C. DeWitt and R. Omnes, p.
161.

\bibitem{eight}
See pp. 172 and 173 of the second reference in Ref.\cite{Hout}.

\bibitem{Schur}
This entire procedure is due to Wigner, and I. Schur applied it earlier to
finite groups.

\bibitem{anal}
This may explain why analytic continuations to Euclidean spaces 
are meaningful for certain calculations.

\bibitem{Kim}
J. J. van der Bij, H. van Dam, and Y. J. Ng, Physica {\bf A116} (1982) 307; 
Y. S. Kim, hep-th/0104051 and references therein.

\bibitem{Tye}
Henry Tye, private communication.

\end{references}
\end{document}